\documentclass[conference,twocolumn,10pt]{IEEEtran} 
\ifCLASSINFOpdf
  \usepackage[pdftex]{graphicx}
  \usepackage{epstopdf} 
\else
\fi
\usepackage{amsmath}
  \usepackage[caption=false,font=footnotesize]{subfig}
\usepackage{url}

\newcommand{\comment}[1]{{}}

\usepackage{amssymb}
\usepackage{xspace}

\newcommand{\complex}{\ensuremath{\mathbb{C}}\xspace}
\newcommand{\complexpow}[2]{\ensuremath{\complex^{#1 \times #2}}\xspace}

\renewcommand{\j}{\ensuremath{\mathrm{j}}}

\usepackage{bm}

\newcommand{\inv}{\ensuremath{^{-1}}\xspace}

\newcommand{\ctrans}{\ensuremath{^{{*}}}\xspace}

\newcommand{\mat}[1]{\ensuremath{\mathbf{#1}}\xspace} 
\renewcommand{\vec}[1]{\ensuremath{\mathbf{#1}}\xspace} 

\newcommand{\pnorm}[2]{\ensuremath{\left\lVert{#2}\right\rVert_{#1}}\xspace}

\newcommand{\entry}[2]{\ensuremath{\left[{#1}\right]_{#2}}\xspace}

\newcommand{\svd}[1]{\ensuremath{\mat{U}_{#1} \ \mat{\Sigma}_{#1} \ \mat{V}\ctrans_{#1} }}

\newcommand{\cgauss}[2]{\ensuremath{\mathcal{N}_{\complex} \left( {#1} , {#2} \right) }\xspace} 
\newcommand{\ev}[1]{\ensuremath{\mathbb{E}\left[{#1}\right]}\xspace}

\newcommand{\prebb}{\ensuremath{\mat{F}_{\textrm{BB}}}\xspace} 

\newcommand{\prerf}{\ensuremath{\mat{F}_{\mathrm{RF}}}\xspace} 
\newcommand{\combb}{\ensuremath{\mat{W}_{\mathrm{BB}}}\xspace} 
\newcommand{\comrf}{\ensuremath{\mat{W}_{\mathrm{RF}}}\xspace} 

\newcommand{\channel}{\ensuremath{\mat{H}}\xspace} 
\newcommand{\channeltilde}{\ensuremath{\tilde{\mat{H}}}\xspace} 
\newcommand{\channeltildedelta}{\ensuremath{\Delta\tilde{\mat{H}}}\xspace} 
\newcommand{\channeltildebar}{\ensuremath{\hat{\tilde{\mat{H}}}}\xspace} 

\newcommand{\Nt}{\ensuremath{N_\mathrm{t}}\xspace} 
\newcommand{\Nr}{\ensuremath{N_\mathrm{r}}\xspace} 

\newcommand{\Nrftx}{\ensuremath{L_{\mathrm{t}}}\xspace} 
\newcommand{\Nrfrx}{\ensuremath{L_{\mathrm{r}}}\xspace} 

\newcommand{\Ns}{\ensuremath{N_\mathrm{s}}\xspace} 

\newcommand{\Pt}{\ensuremath{P_{\mathrm{tx}}\xspace}}

\newcommand{\symbvec}{\ensuremath{\vec{s}}\xspace}
\newcommand{\noisevec}{\ensuremath{\vec{n}}\xspace}

\newcommand{\numrays}{\ensuremath{N_{\mathrm{rays}}}\xspace}
\newcommand{\numclust}{\ensuremath{N_{\mathrm{clust}}}\xspace}

\newcommand{\frobenius}[1]{\ensuremath{\pnorm{\textsf{F}}{#1}}}
\newcommand{\frobeniustwo}[1]{\ensuremath{||#1||^{2}_{\textsf{F}}}}

\newcommand{\aoa}{\ensuremath{\theta}\xspace}
\newcommand{\aod}{\ensuremath{\phi}\xspace}

\newcommand{\rxresponsevector}{\ensuremath{\vec{a}_{\mathrm{r}}(\azimuth_{i,j},\elevation_{i,j})}\xspace}
\newcommand{\txresponsevector}{\ensuremath{\vec{a}_{\mathrm{t}}(\azimuth_{i,j},\elevation_{i,j})}\xspace}

\renewcommand{\rxresponsevector}{\ensuremath{\vec{a}_{\mathrm{r}}}\xspace}
\renewcommand{\txresponsevector}{\ensuremath{\vec{a}_{\mathrm{t}}}\xspace}

\newcommand{\snr}{\ensuremath{\mathrm{SNR}}\xspace}

\newcommand{\node}[1]{\unskip\ensuremath{^{^{\left(#1\right)}}}\xspace}

\newcommand{\ctransnode}[1]{\ensuremath{^{^{\left(#1\right){*}}}}\xspace}

\usepackage{xspace}
\usepackage[acronym,nogroupskip,nonumberlist,nopostdot]{glossaries}
\makeglossaries

\usepackage{xcolor}

\newacronym{snr}{SNR}{signal-to-noise ratio}
\newacronym{sinr}{SINR}{signal-to-interference-plus-noise ratio}
\newacronym{inr}{INR}{interference-to-noise ratio}
\newacronym{sir}{SIR}{signal-to-interference ratio}
\newacronym{ian}{IAN}{interference as noise}
\newacronym{ber}{BER}{bit error rate}
\newacronym{pn}{PN}{pseudorandom noise}
\newacronym{bfsk}{BFSK}{binary frequency shift keying}
\newacronym{fh}{FH}{frequency-hopped}
\newacronym{fh-bfsk}{FH-BFSK}{frequency-hopped binary frequency shift keying}
\newacronym{crc}{CRC}{cyclic redundancy check}
\newacronym{isi}{ISI}{intersymbol interference}
\newacronym{dsss}{DSSS}{direct-sequence spread spectrum}
\newacronym{ofdm}{OFDM}{orthogonal frequency-division multiplexing}
\newacronym{ofdma}{OFDMA}{orthogonal frequency-division multiple access}
\newacronym{sdr}{SDR}{software-defined radio}
\newacronym{tx}{TX}{transmitter}
\newacronym{rx}{RX}{receiver}
\newacronym{fdd}{FDD}{frequency-division duplexing}
\newacronym{tdd}{TDD}{time-division duplexing}
\newacronym{fdma}{FDMA}{frequency-division multiple access}
\newacronym{tdma}{TDMA}{time-division multiple access}
\newacronym{sdma}{SDMA}{space-division multiple access}
\newacronym[plural=MPCs]{mpc}{MPC}{multipath component}
\newacronym{mui}{MUI}{multi-user interference}

\newacronym{ls}{LS}{least-squares}
\newacronym{lms}{LMS}{least mean squares}
\newacronym{rls}{RLS}{recursive least-squares}
\newacronym{rzf}{RZF}{regularized zero-forcing}
\newacronym{mmse}{MMSE}{minimum mean square error}
\newacronym{lmmse}{LMMSE}{linear minimum mean square error}
\newacronym{mse}{MSE}{mean square error}
\newacronym{fft}{FFT}{fast Fourier transform}
\newacronym{dft}{DFT}{discrete Fourier transform}
\newacronym{dtft}{DTFT}{discrete-time Fourier transform}
\newacronym{ctft}{CTFT}{continuous-time Fourier transform}
\newacronym{ml}{ML}{machine learning}
\newacronym[plural=NNs]{nn}{NN}{neural network}
\newacronym[plural=RNNs]{rnn}{RNN}{recurrent neural network}
\newacronym[plural=ADCs]{adc}{ADC}{analog-to-digital converter}
\newacronym[plural=DACs]{dac}{DAC}{digital-to-analog converter}
\newacronym[plural=FPGAs]{fpga}{FPGA}{field-programmable gate array}
\newacronym{evm}{EVM}{error vector magnitude}
\newacronym{psd}{PSD}{power spectral density}
\newacronym{enob}{ENOB}{effective number of bits}
\newacronym{zf}{ZF}{zero-forcing}
\newacronym{rv}{r.v.}{random variable}
\newacronym{omp}{OMP}{orthogonal matching pursuit}
\newacronym{svd}{SVD}{singular value decomposition}

\newacronym{agc}{AGC}{automatic gain control}
\newacronym{rf}{RF}{radio frequency}
\newacronym{los}{LOS}{line-of-sight}
\newacronym{nlos}{NLOS}{non-line-of-sight}
\newacronym{ple}{PLE}{path loss exponent}
\newacronym[plural=dB]{db}{dB}{decibel}
\newacronym{pa}{PA}{power amplifier}
\newacronym{lna}{LNA}{low noise amplifier}
\newacronym{cw}{CW}{continuous wave}
\newacronym{papr}{PAPR}{peak-to-average power ratio}
\newacronym{usrp}{USRP}{Universal Software Radio Peripheral}
\newacronym{irr}{IRR}{image rejection ratio}
\newacronym{lo}{LO}{local oscillator}
\newacronym{vm}{VM}{vector modulator}
\newacronym{mmwave}{mmWave}{millimeter-wave}

\newacronym{csma}{CSMA}{carrier-sense multiple access}
\newacronym{csmaca}{CSMA/CA}{carrier-sense multiple access with collision avoidance}
\newacronym{csmacd}{CSMA/CD}{carrier-sense multiple access with collision detection}
\newacronym{mac}{MAC}{medium access control}
\newacronym{phy}{PHY}{physical layer}
\newacronym{4g}{4G}{fourth generation}
\newacronym{lte}{LTE}{Long-Term Evolution}
\newacronym{4glte}{4G LTE}{\gls{4g} \gls{lte}}
\newacronym{5g}{5G}{fifth generation}
\newacronym{nr}{NR}{New Radio}
\newacronym{5gnr}{5G NR}{5G New Radio}
\newacronym{ieee}{IEEE}{Institute of Electrical and Electronics Engineers}
\newacronym{wifi}{Wi-Fi}{IEEE 802.11}
\newacronym{lan}{LAN}{local area network}
\newacronym{wlan}{WLAN}{wireless local area network}
\newacronym[plural=BSs]{bs}{BS}{base station}
\newacronym[plural=SBSs]{sbs}{SBS}{small-cell base station}
\newacronym[plural=FD-SBSs]{fdsbs}{FD-SBS}{\gls{fd}-enabled \gls{sbs}}
\newacronym[plural=MBSs]{mbs}{MBS}{macrocell base station}
\newacronym[plural=UEs]{ue}{UE}{user equipment}
\newacronym{ul}{UL}{uplink}
\newacronym{dl}{DL}{downlink}
\newacronym{qos}{QoS}{Quality of Service}
\newacronym{fcc}{FCC}{Federal Communications Commission}
\newacronym{iab}{IAB}{integrated access and backhaul}
\newacronym{fab}{FAB}{fixed access and backhaul}
\newacronym{hetnet}{HetNet}{heterogeneous network}

\newacronym{siso}{SISO}{single-input single-output}
\newacronym{mimo}{MIMO}{multiple-input multiple-output}
\newacronym{sumimo}{SU-MIMO}{single-user \gls{mimo}}
\newacronym{mumimo}{MU-MIMO}{multi-user \gls{mimo}}
\newacronym{bf}{BF}{beamforming}
\newacronym{ca}{CA}{constant amplitude}
\newacronym{ula}{ULA}{uniform linear array}
\newacronym{aoa}{AoA}{angle of arrival}
\newacronym{aod}{AoD}{angle of departure}
\newacronym{dof}{DoF}{degrees of freedom}
\newacronym{csi}{CSI}{channel state information}
\newacronym{csit}{CSIT}{\gls{csi} at the transmitter}
\newacronym{csir}{CSIR}{\gls{csi} at the receiver}

\newacronym{fd}{FD}{in-band full-duplex}
\newacronym{hd}{HD}{half-duplex}
\newacronym{si}{SI}{self-interference}
\newacronym{sic}{SIC}{self-interference cancellation}
\newacronym{soi}{SoI}{signal of interest}
\newacronym{asic}{A-SIC}{analog \acrlong{sic}}
\newacronym{dsic}{D-SIC}{digital \gls{sic}}
\newacronym{star}{STAR}{simultaneous transmit and receive}
\newacronym{warp}{WARP}{Wireless Open-Access Research Platform}
\newacronym{bfc}{BFC}{beamforming cancellation}
\newacronym{ipi}{IPI}{inter-panel-interference}
\newacronym{ipic}{IPIC}{inter-panel-interference cancellation}

\newacronym{qcqp}{QCQP}{quadratically-constrained quadratic programming}
\newacronym{cdf}{CDF}{cumulative density function}

\newacronym{elf}{ELF}{extremely low frequency}
\newacronym{slf}{SLF}{super low frequency}
\newacronym{ulf}{ULF}{ultra low frequency}
\newacronym{vlf}{VLF}{very low frequency}
\newacronym{lf}{LF}{low frequency}
\newacronym{mf}{MF}{medium frequency}
\newacronym{hf}{HF}{high frequency}
\newacronym{vhf}{VHF}{very high frequency}
\newacronym{uhf}{UHF}{ultra high frequency}
\newacronym{shf}{SHF}{super high frequency}
\newacronym{ehf}{EHF}{extremely high frequency}
\newacronym{thf}{THF}{tremendously high frequency}

\newacronym{wncg}{WNCG}{Wireless Networking and Communications Group}
\newacronym{linc}{LINC}{Laboratory of Informatics, Networks, and Communications}
\newacronym{ut}{UT Austin}{The University of Texas at Austin}
\newacronym{uiuc}{UIUC}{University of Illinois at Urbana-Champaign}
\newacronym{usc}{USC}{University of Southern California}
\newacronym{mit}{MIT}{Massachusetts Institute of Technology}
\newacronym{berkeley}{UC Berkeley}{University of California, Berkeley}
\newacronym{osu}{OSU}{Ohio State University}

\newcommand{\mmwave}{\gls{mmwave}\xspace}
\newcommand{\mimo}{\gls{mimo}\xspace}

\newcommand{\rf}{\gls{rf}\xspace}

\newcommand{\sic}{\gls{sic}\xspace}

\newcommand{\asic}{\gls{asic}\xspace}

\newcommand{\figref}[1]{\figurename~\ref{#1}}

\newcounter{mytempeqncnt}

\begin{document}

%
\title{Equipping Millimeter-Wave Full-Duplex with Analog Self-Interference Cancellation}
%
%
%

\author{Ian~P.~Roberts, Hardik~B.~Jain, and Sriram~Vishwanath\\GenXComm, Inc.\\ ianroberts@genxcomminc.com}

\maketitle

\begin{abstract}
There have been recent works on enabling in-band full-duplex operation using \mmwave transceivers. These works are based solely on creating sufficient isolation between a transceiver's transmitter and receiver via \mimo precoding and combining. In this work, we propose supplementing these beamforming strategies with \asic. By leveraging \asic, a portion of the self-interference is cancelled without the need for beamforming, allowing for more optimal beamforming strategies to be used in serving users. We use simulation to demonstrate that even with finite resolution \asic solutions, there are significant gains to be had in sum spectral efficiency. With a single bit of \asic resolution, improvements over a beamforming-only design are present. With 8 bits of \asic resolution, our design nearly approaches that of ideal full-duplex operation. To the best of our knowledge, this is the first \mmwave full-duplex design that combines both beamforming and \asic to achieve simultaneous transmission and reception in-band.
\end{abstract}

%
\IEEEpeerreviewmaketitle

\glsresetall

\section{Introduction} \label{sec:introduction}

Modern wireless networks have turned to the wide bandwidths offered at \mmwave frequencies to satisfy the ever-increasing consumption of information \cite{Andrews_Buzzi_Choi_Hanly_Lozano_Soong_Zhang_2014}. To further capitalize on these wide bandwidths, recent works have proposed designs enabling in-band full-duplex operation at \mmwave \cite{ipr-bfc-2019-globecom,Satyanarayana_El-Hajjar_Kuo_Mourad_Hanzo_2019}. Thus far, the proposed techniques for achieving simultaneous transmission and reception in-band at \mmwave have been by means of \mimo precoding and combining to mitigate the self-interference that would otherwise be incurred. In short, these beamforming methods seek to avoid the \mimo channel between the transmit and receive arrays of a full-duplex \mmwave transceiver.

Significant work on achieving full-duplex capability at sub-$6$ GHz has taken place over the past decade \cite{microsoft_research,stanford_achieving,stanford_practical,stanford_full_duplex_radios}. These works were motivated largely by the fact that operating in a full-duplex fashion doubles the capacity of a communication channel as compared to conventional half-duplex operation. This is due to the fact that, in full-duplex, the entire time-frequency resource is being used by both transmission and reception, rather than being divided between the two. In addition to capacity gains, medium access and latency improvements can be had with full-duplex.

When attempting to transmit while receiving over the same frequencies, a full-duplex transceiver suffers from self-interference---the undesired leakage from a device's transmitter to its own receiver. Unless mitigated sufficiently, the self-interference makes successful reception of a desired signal virtually impossible or severely degraded at best. Methods of \sic take advantage of the fact that a full-duplex transceiver is privy to its own transmit signal, allowing the device to cancel the self-interference with an inverted copy of itself. Fundamental hardware factors have popularized two-stage designs that seek to cancel the incurred self-interference in the analog (\rf) domain followed by in the digital domain. 

While tremendous strides have been made in full-duplex research, almost all of this work has been in application to sub-$6$ GHz communication systems. Moreover, the developed methods do not directly translate well to \mmwave systems. This is largely due to the numerous antennas, wide bandwidths, and high nonlinearity present in \mmwave systems. For this reason, it has been regarded that these existing approaches will not be suitable for achieving \mmwave full-duplex \cite{ipr-bfc-2019-globecom,Satyanarayana_El-Hajjar_Kuo_Mourad_Hanzo_2019}. 


In this paper, however, we propose a design that enables \mmwave full-duplex by combining strategic beamforming and \asic.
While it has been shown that strategic beamforming alone can enable \mmwave full-duplex \cite{ipr-bfc-2019-globecom,ipr-fsbfc-2020-icc}, we demonstrate that a practical \asic solution can offer even further spectral efficiency gains over conventional half-duplex operation. Furthermore, our \asic design does not necessarily have to grow in proportion to the number of antennas but rather with the number of \rf chains---a much smaller quantity in \mmwave systems.
We simulate our proposed system design to validate this.

\textit{Notation:}
We use bold uppercase, \mat{A}, to represent matrices. 
We use bold lowercase, \vec{a}, to represent column vectors. 
We use $(\cdot)$\ctrans, \frobenius{\cdot}, and $\ev{\cdot}$ to represent conjugate transpose, Frobenius norm, and expectation, respectively. 
We use $\left[\mat{A}\right]_{i,j}$ to denote the element in the $i$th row and $j$th column of \mat{A}. 
We use $\left[\mat{A}\right]_{i,:}$ and $\left[\mat{A}\right]_{:,j}$ to denote the $i$th row and $j$th column of \mat{A}, respectively. 
We use $\cgauss{\vec{m}}{\mat{R}}$ as a multivariate circularly symmetric complex Normal distribution with mean $\vec{m}$ and covariance matrix $\mat{R}$.

\section{System Model} \label{sec:system-model}

\begin{figure*}[!t]
	\normalsize
	\centering
	\includegraphics[width=0.85\linewidth]{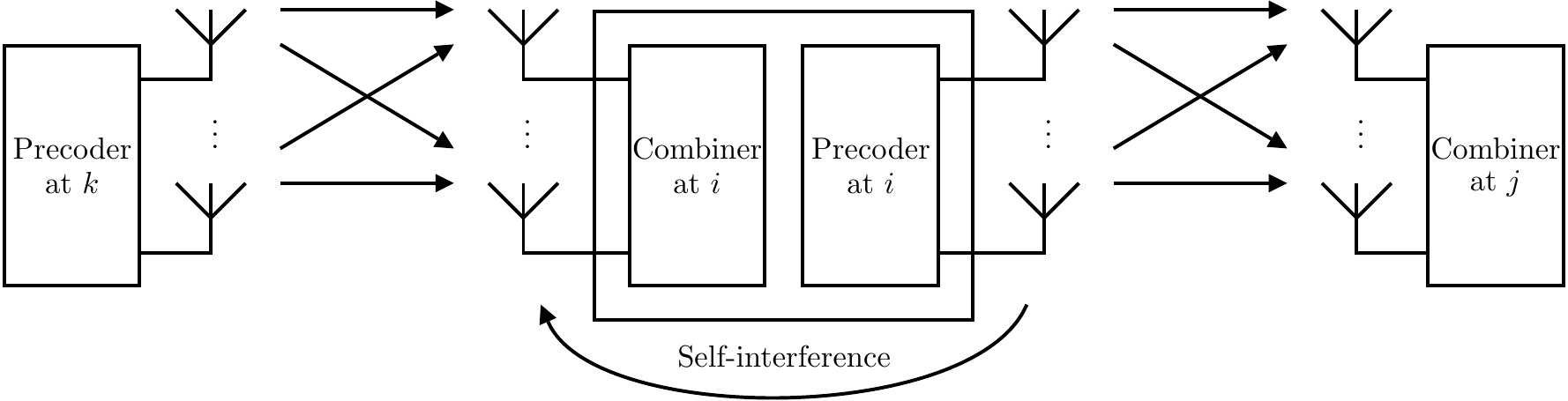}
	\caption{A full-duplex \mmwave device $i$ transmitting to $j$ as it receives from $k$ in-band. In doing so, a \mimo self-interference channel is introduced between the transmit array and the receive array at $i$. Devices $j$ and $k$ can be conventional half-duplex devices or can comprise a single full-duplex device.}
	\label{fig:network}
	\vspace*{4pt}
\end{figure*}

The proposed design in this work enables a \mmwave transceiver $i$ to transmit to a device $j$ while receiving from a device $k$ in-band, as exhibited in \figref{fig:network}. Without a proper design, this sort of full-duplex operation would conventionally be impossible due to the overwhelming received self-interference at $i$. In Section~\ref{sec:contribution}, we present a means to enable such operation. We first introduce our system model, to which our design will be tailored.

Ubiquitous among practical \mmwave systems is the use of hybrid digital/analog beamforming architectures where precoding (or combining) is implemented by the combination of baseband processing and \rf processing---such methods can offer performance comparable to fully-digital beamforming with a reduced number of \rf chains \cite{Heath_Gonzalez-Prelcic_Rangan_Roh_Sayeed_2016}. We assume devices $i$, $j$, and $k$ all employ hybrid beamforming. Specifically, we consider the case of fully-connected hybrid beamforming whose \rf beamformers only have phase control. 

For a device $m \in \{i,j,k\}$, we use the following notation. 
Let $\Nt\node{m}$ and $\Nr\node{m}$ be the number of transmit and receive antennas, respectively.
Let $\prebb\node{m}$ be the baseband precoder and $\prerf\node{m}$ be the \rf precoder, responsible for transmitting from $m$. 
Let $\combb\node{m}$ be the baseband combiner and $\comrf\node{m}$ be the \rf combiner, responsible for receiving at $m$.
Connecting the baseband and \rf stages, let $\Nrftx\node{m}$ be the number of transmit \rf chains and $\Nrfrx\node{m}$ be the number of receive \rf chains.

We assume devices $i$ and $j$ and devices $i$ and $k$ are separated in a far-field fashion. We model the channels $\channel_{ij}$ (from $i$ to $j$) and $\channel_{ki}$ (from $k$ to $i$) with the Saleh-Valenzuela representation where propagation from one device to another is modeled by clusters of rays \cite{Heath_Gonzalez-Prelcic_Rangan_Roh_Sayeed_2016}. Explicitly, channels $\channel_{ij}$ and $\channel_{ki}$ are modeled as follows, where $m,n \in \{i,j,k\}$, 
\begin{align}
\channel_{mn} = \sqrt{\frac{\Nt\node{m} \Nr\node{n}}{\numrays \numclust}} \sum_{u=1}^{\numclust} \sum_{v=1}^{\numrays} \beta_{uv} \rxresponsevector(\aoa_{uv}) \txresponsevector\ctrans(\aod_{uv}). \label{eq:desired-channel}
\end{align}
In each channel, $\numrays$ and $\numclust$ are independent random variables dictating the number of rays per cluster and number of clusters, respectively. The complex gain of ray $v$ from cluster $u$ is given as $\beta_{uv} \sim \cgauss{0}{1}$. A ray's \gls{aod} and \gls{aoa} are given as $\aod_{uv}$ and $\aoa_{uv}$, respectively. The transmit and receive array responses at these angles are given as $\rxresponsevector(\aoa_{uv})$ and $\txresponsevector(\aod_{uv})$, respectively.

To model the channel between the transmit and receive arrays of device $i$, we use the following summation with Rician factor $\kappa$ \cite{spherical-wave-mimo,Satyanarayana_El-Hajjar_Kuo_Mourad_Hanzo_2019}.
\begin{align} \label{eq:si-channel-rice}
	\channel_{ii} &= \sqrt{\frac{\kappa}{\kappa + 1}} \channel^{\textrm{LOS}}_{ii} + \sqrt{\frac{1}{\kappa + 1}} \channel^{\textrm{NLOS}}_{ii}
\end{align}
The \gls{los} component is described in a near-field (spherical-wave) fashion as
\begin{align} \label{eq:si-channel-los}
	\left[\channel^{\textrm{LOS}}_{ii}\right]_{u,v} &= \frac{\rho}{r_{u,v}}\exp \left(-\j 2 \pi \frac{r_{u,v}}{\lambda} \right)
\end{align}
where $r_{u,v}$ is the distance between the $u$th transmit antenna and the $v$th receive antenna, $\lambda$ is the carrier wavelength, and $\rho$ ensures that the channel is normalized such that $\ev{\frobeniustwo{\channel_{ii}}} = \Nt\node{i}\Nr\node{i}$.
The \gls{nlos} component is modeled in a far-field fashion using \eqref{eq:desired-channel} .
It is worthwhile to remark that the self-interference channel is not yet well-characterized for \mmwave systems, meaning this model may not align well with practice. However, we expect our design herein will translate well to more practically-sound self-interference channels, for we do not rely on its specific structure or properties.

We assume that the large scale power gain between two devices $m,n\in\{i,j,k\}$ is given by $G^2_{mn}$. Furthermore, we assume a device $m\in\{i,j,k\}$ transmits with a total power of $\Pt\node{m}$.
Let $\noisevec\node{m} \sim \cgauss{\mat{0}}{\sigma^2\mat{I}}$ be the $\Nr\node{m} \times 1$ additive noise vector incurred at the receive array of $m$, where $\sigma^2$ represents a per-antenna noise variance (common across devices).

To establish some power normalizations, we make the following declarations for device $m\in\{i,j,k\}$. Let $\Ns\node{m}$ be the number of data streams transmitted to device $m$. Let $\symbvec\node{m}$ be the $\Ns\node{m} \times 1$ symbol vector intended for device $m$, where 
\begin{align}
\ev{\symbvec\node{m}\symbvec\ctransnode{m}} = \frac{1}{\Ns\node{m}} \ \mat{I}.
\end{align}
Finally, we impose unit power allocation across streams. (While generally suboptimal, we make this declaration for simplicity.) To enforce this, we normalize each stream's effective precoding vector such that
\begin{align}
{\Big\Vert\prerf\node{m}\entry{\prebb\node{m}}{:,\ell}\Big\Vert_{\textsf{F}}}^2 = 1 \ \forall \ \ell \in [0,\Ns\node{m}-1]  \label{eq:power}
\end{align}
which ensures that, for all $m \in \{i,j,k\}$,
\begin{align}
\frobenius{\prerf\node{m}{\prebb\node{m}}}^2 = \Ns\node{m}.
\end{align}

\begin{figure*}[!t]
	\normalsize
	\centering
	\includegraphics[width=\linewidth]{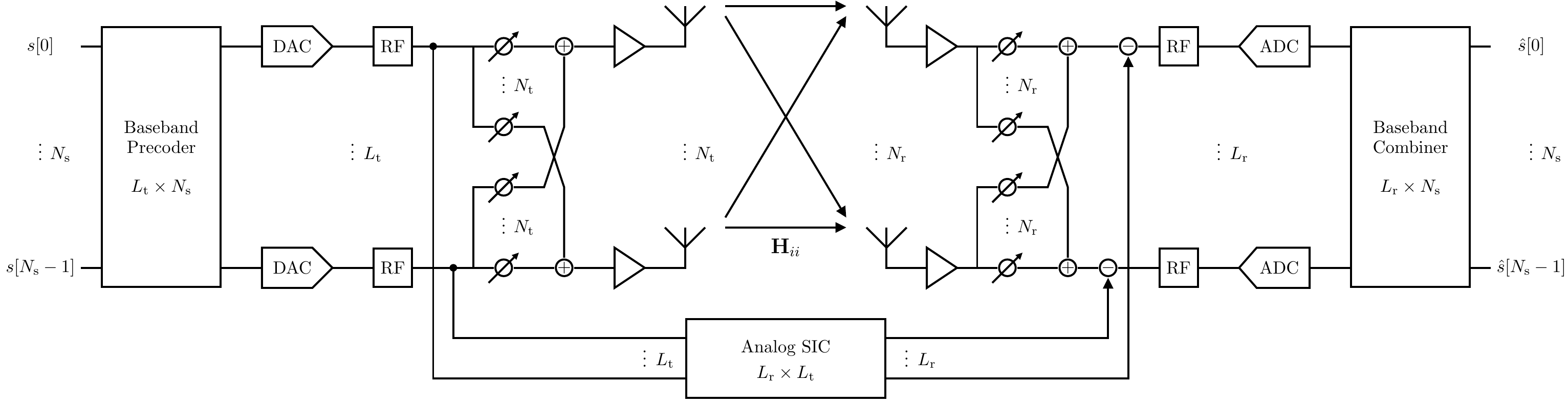}
	\caption{A block diagram describing our full-duplex \mmwave transceiver architecture utilizing \asic in conjunction with hybrid beamforming. A portion of the transmit signal is tapped off before the \rf precoder at $i$. The \asic filter is effectively an $\Nrfrx\node{i}\times\Nrftx\node{i}$ matrix of complex weights. The output of the \asic is injected after the \rf combiner at the receiver of $i$.}
	\label{fig:transceiver}
	\vspace*{4pt}
\end{figure*}
We define a \gls{snr} quantity between two devices $m,n\in\{i,j,k\}$ as
\begin{align}
\snr_{mn} \triangleq \frac{\Pt\node{m} G^2_{mn}}{\sigma^2}.
\end{align}

\section{Proposed Design} \label{sec:contribution}

We now present a solution to enable our system to operate in a full-duplex fashion. Our design consists of two modes of self-interference mitigation: ($i$) using \asic at device $i$ to cancel a portion of the self-interference and ($ii$) using strategic beamforming to mitigate the residual self-interference following \asic. The design presented herein begins with beamtraining (Stage 0), followed by configuring the \asic (Stage 1), and finally beamforming (Stage 2).

\subsection*{Stage 0: Beamtraining}
Initial access and channel estimation are challenging at \mmwave due to the poor coverage, high path loss, and hybrid beamforming architecture. While there have been many sophisticated works addressing these challenges, practical systems have turned to methods of ``beamtraining'', whereby a sort of search allows a device to establish a link with another without the need for prior channel knowledge. Beamtraining schemes can be found in \gls{5gnr} and IEEE 802.11ad.
In general, beamtraining between two devices consists of a codebook based search through \rf beamformers at each device. For each transmit-receive pair of \rf beamformers, the received signal strength is measured. The choice of \rf beamformers can then be chosen based on the pairs that offer sufficient signal strength (e.g., the strongest pairs).

For our design, we do not rely on a particular beamtraining strategy. We simply assume one has taken place.
Following beamtraining between a transmitting device and receiving device, the \rf precoding matrix and the \rf combining matrix are fixed for the channel between the two. 
We assume that beamtraining has taken place for communication from $i$ to $j$ and from $k$ to $i$. The beamtraining period provides selections for $\prerf\node{i}$ and $\comrf\node{j}$ as well as for $\prerf\node{k}$ and $\comrf\node{i}$. With the \rf beamformers on all links fixed, communication effectively reduces to fully-digital \mimo, where the \glspl{dac} and \glspl{adc} have direct access to their \textit{effective channels} as seen through the \rf beamformers. These effective channels are written as 
\begin{align}
\channeltilde_{ij} &\triangleq \comrf\ctransnode{j} \channel_{ij} \prerf\node{i} \label{eq:effective-rf-channel-ij} \\
\channeltilde_{ki} &\triangleq \comrf\ctransnode{i} \channel_{ki} \prerf\node{k} \label{eq:effective-rf-channel-ki} \\
\channeltilde_{ii} &\triangleq \comrf\ctransnode{i} \channel_{ii} \prerf\node{i} \label{eq:effective-rf-channel-ii}
\end{align}
where \eqref{eq:effective-rf-channel-ij} and \eqref{eq:effective-rf-channel-ki} are the effective desired channels and \eqref{eq:effective-rf-channel-ii} is the effective self-interference channel.
We remark that channel estimation now becomes not only more straightforward but also much more reduced since channels that were once of size based on the number of antennas are now based merely on the number of \rf chains---a significant reduction in \mmwave communication systems. In the design that follows, we assume perfect channel estimation of these effective channels along with full \gls{csir} and \gls{csit}. We also assume perfect estimation of the link \glspl{snr}.

\subsection*{Stage 1: Analog SIC Design}
We now focus on using \asic to mitigate a portion of the self-interference. Comparing \mmwave to sub-$6$ GHz systems, the self-interference channel $\channel_{ii}$ has grown to be quite large---of size $\Nr\node{i} \times \Nt\node{i}$. This introduces challenges in creating \asic solutions that are not prohibitive in size, cost, power consumption, or complexity. However, with beamtraining having taken place, we can consider the effective self-interference channel $\channeltilde_{ii}$, which is of size $\Nrfrx\node{i} \times \Nrftx\node{i}$ (e.g., $\channel_{ii} \in \complexpow{64}{64}$ while $\channeltilde_{ii} \in \complexpow{2}{2}$).

To further describe the goal of \asic, consider \figref{fig:transceiver} which depicts the full-duplex transceiver $i$ whose transmitter and receiver are linked by the self-interference channel matrix $\channel_{ii}$. Note that the \asic taps off of the transmit signal before the \rf precoder and is injected (by subtraction) following the \rf combiner. The goal of \asic is to replicate the effective self-interference channel matrix $\channeltilde_{ii}$ as accurately as possible. If done properly, this replicated self-interference will be subtracted from the true self-interference leaving a weak residual self-interference channel. In other words, we seek to use \asic to create some matrix $\channeltildebar_{ii}$ that is close to $\channeltilde_{ii}$. (It also must account for proper scaling via knowledge of $\snr_{ii}$.)

We assume that our \asic solution is limited to some finite resolution. For example, it could be the case that the \asic solution is implemented using digitally-controlled hardware (e.g., stepped attenuators, phase shifters) or is configured via digital hardware.
Let us assume that each entry in $\channeltildebar_{ii}$ can only be expressed using $M$ bits of resolution. Even with the assumption that $\channeltilde_{ii}$ is known, the resolution of \asic will introduce residual (quantization) error when attempting to replicate it. Put simply, $\channeltilde_{ii}$ can be written as
\begin{align}
\channeltilde_{ii} = \channeltildebar_{ii} + \channeltildedelta_{ii} \label{eq:asic-error}
\end{align}
where $\channeltildedelta_{ii}$ captures the error in \asic's attempt to replicate $\channeltilde_{ii}$ due to quantization. Stage 2 of our full-duplex design will handle the residual error $\channeltildedelta_{ii}$ by strategically beamforming to avoid it.

\subsubsection*{Remarks}
It is important to note that the \asic design that we have presented relies heavily on the transceiver architecture depicted in \figref{fig:transceiver} along with a few assumptions. First, since the input signal to the \asic is tapped off before the transmitter \glspl{pa}, it will not capture the nonlinearities that they may introduce. One can either assume that the \glspl{pa} are sufficiently linear or that the nonlinearities are dealt with using further digital processing. Secondly, since the output signal of the \asic is combined after the receive \glspl{lna}, it is implicitly assumed that the self-interference strength is not saturating the \glspl{lna}. This assumption can be justified with sufficiently linear \glspl{lna} or with sufficient isolation between the transmit and receive arrays at $i$ (i.e., $G_{ii}$). Finally, we point out that the need for \asic is driven by the fact that the receive \glspl{adc} have finite resolution (i.e., finite dynamic range). Given our previous assumptions, having infinite dynamic range \glspl{adc} would allow all of the \sic processing to take place in the digital domain. With a finite dynamic range, however, the need for \asic is immediate: sufficient cancellation must take place before the \glspl{adc} to ensure the self-interference does not saturate the \glspl{adc}, reducing a desired receive signal's effective resolution. We assume this is the case when \gls{asic} is in use. We would like to remark that the model in \eqref{eq:asic-error} could also represent other sorts of error in \asic such as configuration errors.

The purpose of this work is not to produce a design that is flushed for truly practical \mmwave systems, but merely to make a stride in that direction. Our ongoing and future work will address more of these concerns to inch our way toward a truly practical \mmwave full-duplex design.

\subsection*{Stage 2: Precoding and Combining Design}
With the \rf precoding and combining matrices fixed from beamtraining, our design will be in tailoring the baseband precoding and combining matrices. The goal is to design the baseband precoders and combiners to maintain communication on the desired links while also mitigating the residual self-interference following \asic. 


There is no closed-form design of linear precoders and combiners to maximize the sum spectral efficiency of our three device scenario. However, there are designs that can offer impressive performance, such as those seen in \cite{ipr-bfc-2019-globecom,ipr-fsbfc-2020-icc}. 

We begin by taking the \gls{svd} of the effective desired channels as
\begin{align}
\channeltilde_{ij} &= \svd{ij} \\
\channeltilde_{ki} &= \svd{ki}
\end{align}
whose singular values decrease along their respective diagonals.
We will receive at $j$ along the strongest $\Ns\node{j}$ left singular vectors of $\channeltilde_{ij}$ and will transmit from $k$ along the $\Ns\node{i}$ strongest right singular vectors of $\channeltilde_{ki}$. 
\begin{align}
\combb\node{j} = \entry{\mat{U}_{ij}}{:,0:\Ns\node{j}-1} \\
\prebb\node{k} = \entry{\mat{V}_{ki}}{:,0:\Ns\node{i}-1}
\end{align}
This is, of course, the conventional route taken for maximizing the so-called Gaussian mutual information (with proper power allocation) \cite{heath_lozano}. Such optimality only applies to the half-duplex sense; in our case, we are plagued by self-interference, requiring us to design for such as follows.

With only the baseband precoder and combiner at the full-duplex device $i$ left to be configured, we take a moment to make a few remarks. We could use either the precoder or the combiner or both to mitigate the self-interference. The precoder could avoid contributing interference, the combiner could avoid receiving interference, or both. Intuitively, it seems the precoder and combiner should share the responsibility of mitigating the self-interference. By mitigating the self-interference with the precoder and combiner, the spectral efficiencies of the two links will degrade since the precoder and combiner will not be along the singular vectors. This is a price we pay for operating in a full-duplex fashion.

\begin{figure*}[!t]
	\normalsize
	\setcounter{mytempeqncnt}{\value{equation}}
	\setcounter{equation}{18}
	\begin{align}
	\prebb\node{i} = \entry{\left(\channel_{\mathrm{des}}\ctrans \channel_{\mathrm{des}} + \frac{\snr_{ii}}{\snr_{ij}} \channel_{\mathrm{int}}\ctrans \channel_{\mathrm{int}} + \frac{\Ns\node{j}}{\snr_{ij}}\mathbf{I} \right)\inv \channel_{\mathrm{des}}\ctrans}{:,0:\Ns\node{j}-1} 	\label{eq:mmse}
	\end{align}
	\setcounter{equation}{\value{mytempeqncnt}}
	\hrulefill
	\vspace*{4pt}
\end{figure*}

We decide to preserve the link with device $k$, meaning the combiner at $i$ will be the optimal half-duplex one (left singular vectors). 
\begin{align}
\combb\node{i} = \entry{\mat{U}_{ki}}{:,0:\Ns\node{i}-1}
\end{align}
This will place the whole responsibility on mitigating the self-interference at the precoder of $i$. To design the precoder so that it avoids contributing self-interference, we will design it in a \gls{mmse} fashion---the goal is to balance the amount of energy pushed into a desired channel versus the amount pushed into an interference-plus-noise channel. 
This is accomplished with the design shown in \eqref{eq:mmse}, where 
\begin{align}
\channel_{\mathrm{des}} &\triangleq \combb\ctransnode{j} \comrf\ctransnode{j} \channel_{ij} \prerf\node{i} \\
\channel_{\mathrm{int}} &\triangleq \combb\ctransnode{i} \underbrace{\left(\comrf\ctransnode{i} \channel_{ii} \prerf\node{i} - \channeltildebar_{ii}\right)}_{=\channeltildedelta_{ii}} 
\end{align}
are the desired and interference channels, respectively, that our \gls{mmse} precoder is concerned with. Courtesy of \asic, note that the interference channel is comprised of the residual $\channeltildedelta_{ii}$ rather than $\channeltilde_{ii}$---this is how beamforming strategies can benefit from the use of \asic. Following proper power normalizations according to \eqref{eq:power}, our precoding and combining design is complete: all \rf beamformers were set during beamtraining and all baseband beamformers were set during this design.

\addtocounter{equation}{1}

\section{Simulation and Results} \label{sec:simulation-results}

To validate our design, we simulated it in a Monte Carlo fashion, varying the \gls{snr} and recording the achieved spectral efficiency on each link.
Half-wavelength \glspl{ula} with $32$ antennas were used at each device's transmitter and receiver. We allocated $2$ \rf chains to each transmitter and receiver and provided an additional $2$ to the transmitter of $i$---this offers more dimensions to suppress the interference. On both links, we transmitted $2$ streams. The \gls{aod} and \gls{aoa} were drawn independently and uniformly on $[0,\pi]$. For each of the two desired channels, the number of rays per cluster and number of clusters were drawn uniformly on $[1,10]$ and $[1,6]$, respectively.

To create the self-interference channel $\channel_{ii}$, we stacked the transmit and receive arrays at $i$ vertically, separated by $10$ wavelengths. We used a carrier frequency of $28$ GHz. The \gls{nlos} portion was created with a number of rays per cluster and a number of clusters drawn uniformly on $[1,6]$ and $[1,3]$, respectively. We used a Rician factor of $\kappa = 20$ dB, indicating that the \gls{los} dominates. We used a fixed $\snr_{ii} = 40$ dB, indicating a moderately strong self-interference strength. For beamtraining, we searched across a \gls{dft} codebook, taking the strongest beam pairs.

\begin{figure}[!t]
	\centering
	\includegraphics[width=\linewidth]{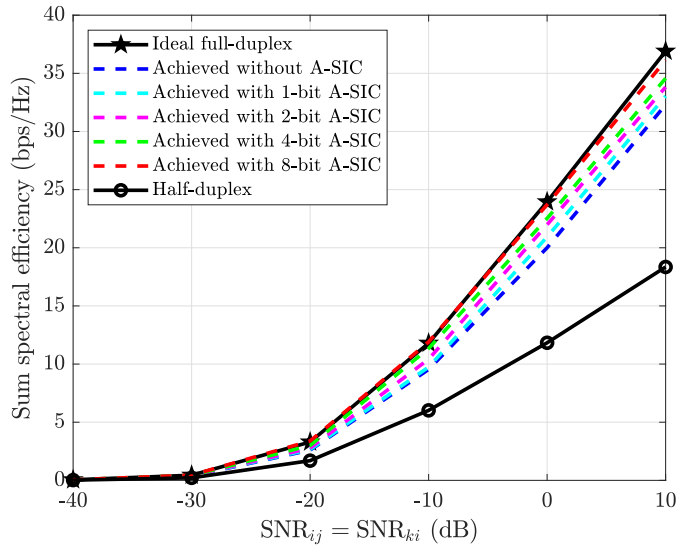}
	\caption{Sum spectral efficiency as a function of \gls{snr} for various scenarios. As the resolution of \asic improves, the sum spectral efficiency approaches that of ideal (interference-free) full-duplex.}
	\label{fig:results}
\end{figure}

The results of our simulations can be seen in \figref{fig:results}. 
For the sake of analysis, we consider the case when the two links are equal in \gls{snr} (i.e., $\snr_{ij} = \snr_{ki}$).
To evaluate our design's performance, we compare the achieved sum spectral efficiency to that of ideal full-duplex---interference-free transmission and reception. In other words, ideal full-duplex is achieved when there is complete isolation between transmission and reception.
When operating in a half-duplex fashion, the rate would simply be half of ideal full-duplex, as indicated in \figref{fig:results}.

Now, let us examine the achieved spectral efficiencies between half-duplex and ideal full-duplex. First, if we do not use \asic at all and rely solely on beamforming (shown in dashed blue), we achieve admirable results that are certainly superior to half-duplex but fall significantly short of ideal full-duplex.
By supplementing our beamforming strategy with \asic, we can see that attractive gains can be had. Even with only a $1$-bit \asic, noticeable spectral efficiency gains are present. As the resolution of \asic improves, it approaches closer and closer to ideal full-duplex. With an $8$-bit \asic, nearly all of the gains offered by full-duplex are obtained. 

Without \asic, beamforming alone can sufficiently mitigate the self-interference while also maintaining service to $j$ and from $k$. With \asic, a portion of the self-interference is mitigated by \asic, allowing us to more optimally beamform on each link. As the resolution of \asic improves, it more accurately cancels the self-interference, leaving a weaker and weaker residual self-interference channel that our beamforming design attempts to avoid. This drives our \gls{mmse} precoder to better serve $j$.

\section{Conclusion} \label{sec:conclusion}



While there has been recent work to enable full-duplex \mmwave systems via beamforming solutions, in this paper we suggest that such systems can be supplemented with \asic. Even with finite resolution \asic solutions, we demonstrate that significant gains can be had over \mmwave full-duplex designs that rely solely on beamforming to mitigate the self-interference. Our design was validated in simulation which suggests that the sum spectral efficiency achieved during full-duplex operation can approach that of an ideal full-duplex system. 
To the best of our knowledge, this is the first work that combines beamforming with \asic to enable simultaneous transmission and reception in-band at \mmwave.


\bibliographystyle{bibtex/IEEEtran}
\bibliography{refs}

\end{document}